\documentclass[prl,twocolumn,amsmath,amssymb,superscriptaddress]{revtex4-1}

\usepackage{graphicx}
\usepackage{dcolumn}
\usepackage{hhline}
\usepackage{bm,color}
\usepackage{float}
\usepackage{amsmath}
\usepackage{amsfonts}
\usepackage{amssymb}
\usepackage{epstopdf}

\begin{document}

\title{Similarities and differences between infinite-layer nickelates and cuprates and implications for superconductivity}

\author{A. S. Botana}
\affiliation{Department of Physics, Arizona State University, Tempe, AZ 85287}
\author{M. R. Norman}
\affiliation{Materials Science and Engineering Division, Argonne National Laboratory, Argonne, IL 60439}
\date{\today}

\begin{abstract}
We have revisited the electronic structure of infinite-layer RNiO$_2$ (R= La, Nd) in light of the recent discovery of superconductivity in Sr-doped NdNiO$_2$.  From a comparison to their cuprate counterpart CaCuO$_2$, we derive
essential facts related to their electronic structures, in particular the values for various hopping parameters and energy splittings, and the influence of the
spacer cation.  From this detailed comparison, we comment on expectations in regards to superconductivity.
In particular, both materials exhibit a large ratio of longer-range hopping to near-neighbor hopping which should be conducive for superconductivity.
\end{abstract}

\maketitle


The quest for finding cuprate analogs in connection with high-T$_c$ superconductivity has followed several routes \cite{norman-RPP}.
Looking at nickelates has been an obvious path: nickel is next to copper in the periodic
table, and if the former could be realized in the 1+ oxidation state,
it would be isoelectronic with Cu$^{2+}$ \cite{anisimov}.
This oxidation state is indeed realized in the infinite-layer square planar
materials RNiO$_2$ (R= La, Nd) \cite{crespin,hayward} with the same P4/mmm crystal structure as that of the parent compound of high-T$_c$ cuprates,
CaCuO$_2$ (Fig.~1 of Ref.~\cite{supplementary}). The latter has a T$_c$ of 110 K upon hole doping \cite{azuma}.
Still, doubts have been raised that RNiO$_2$ would be cuprate analogs.
Available transport data indicate that LaNiO$_2$ is not a charge transfer insulator \cite{ikeda,ikeda2} and there is no experimental evidence for antiferromagnetic order in any RNiO$_2$ material
\cite{hayward_nd}. Electronic structure calculations of LaNiO$_2$ indicate significant differences from CaCuO$_2$ due to the presence of low lying La-5$d$ states, as well as an increased splitting between
Ni-$d$ and O-$p$ levels \cite{pickett,liu,antia-Pd}.

The recent observation of superconductivity in Sr-doped NdNiO$_2$, though, begs a reconsideration of this earlier thinking \cite{new}.
In this new context, we reanalyze the electronic structure of RNiO$_2$ 
and do a detailed comparison to that of CaCuO$_2$. We find that the important
$pd$ and $pp$ hopping energies are comparable in the Ni and Cu cases, with a large $t^\prime/t$ ratio. This has been shown by Pavarini {\it et al.} to correlate with a high T$_c$ in the cuprates
\cite{pavarini}.
The splitting of the two $e_g$ orbital
energies is similar, which is also thought to be relevant for T$_c$ \cite{sakakibara}.
But the difference in charge-transfer 
energies $\Delta$= $\epsilon_d-\epsilon_p$ is significant, being much larger in the Ni case.  This puts RNiO$_2$ outside the bounds for superconductivity according to the considerations
of Weber {\it et al.}~\cite{weber}.  This indicates the need to reexamine this criterion in light of the observation of superconductivity, though perhaps the low value of T$_c$ \cite{new} is due to the increased $\Delta$.
Moreover, the large hole-like Fermi surface associated with the $d_{x^2-y^2}$ state is self-doped in the Ni case due
to two small electron pockets of 5$d$ origin.
This is consistent with transport data, which indicates weak localization for
RNiO$_2$ \cite{ikeda2,new} similar to what is seen in underdoped (as opposed to undoped) cuprates.
This also implies that the value for optimal doping could differ from that of the cuprates.

\textit{Computational Methods.} Electronic structure calculations were performed using the all-electron, full potential
code WIEN2k \cite{wien2k} based on the augmented plane wave
plus local orbitals (APW + lo) basis set.
The Perdew-Burke-Ernzerhof version of the generalized
gradient approximation (GGA) \cite{pbe} was used for the non-magnetic calculations. The missing correlations beyond GGA at Ni sites were taken into account through LDA+U calculations. Two LDA+$U$ schemes were used: the `fully localized limit' (FLL) and the `around mean field' (AMF) \cite{sic, amf}. For both schemes, we have studied the evolution of the electronic structure with increasing $U$  ($U_{Ni}$= 1.4 to 6 eV, $J$= 0.8 eV). The lattice parameters used for LaNiO$_2$ were $a$= 3.96 \AA, $c$= 3.37 \AA, for NdNiO$_2$ $a$= 3.92 \AA, $c$= 3.28 \AA, for CaCuO$_2$  $a$= 3.86 \AA, $c$= 3.20 \AA. Supercells of size 2$\times$2, and 3$\times$3 relative to the primitive P4/mmm
cell were employed to study the effect of Sr doping. 

To look for possible magnetic solutions, $\sqrt{2}$$\times$$\sqrt{2}$ and $\sqrt{2}$$\times$$\sqrt{2}$$\times$2 cells were constructed. Calculations for different magnetic configurations were performed: (i) ferromagnetic (FM), (ii) antiferromagnetic (AFM) in plane with FM coupling out of plane, (iii) AFM in plane with AFM coupling out of plane.
For all calculations, we converged using R$_{mt}$K$_{max}$ = 7.0.
The muffin-tin radii used were typical values of 2.5 \AA~for La and Nd, 2.35 \AA~for Ca, 2 \AA~for Ni, 1.95 \AA~for Cu and 1.72 \AA~for O. A dense mesh of 25$\times$25$\times$25 $k$-points in the irreducible Brillouin zone was used for the non-magnetic calculations.

To further understand the electronic structure and the comparison of Ni to Cu, we performed an analysis based on maximally localized Wannier functions
(MLWFs) \cite{mlwf}. For the spread
functional minimization, we used WANNIER90 \cite{wannier90}. Post-processing of MLWFs to generate tight-binding band structures, hopping integrals, and plots of Wannier orbitals were done with WIEN2WANNIER \cite{wien2wannier}.  These values were also used as start values for a Slater-Koster fit of the electronic structure \cite{slater}.
In addition, we performed a simple tight-binding fit of the dominant $d_{x^2-y^2} - p\sigma$ antibonding band at the Fermi energy.

\begin{figure}
\includegraphics[width=0.85\columnwidth]{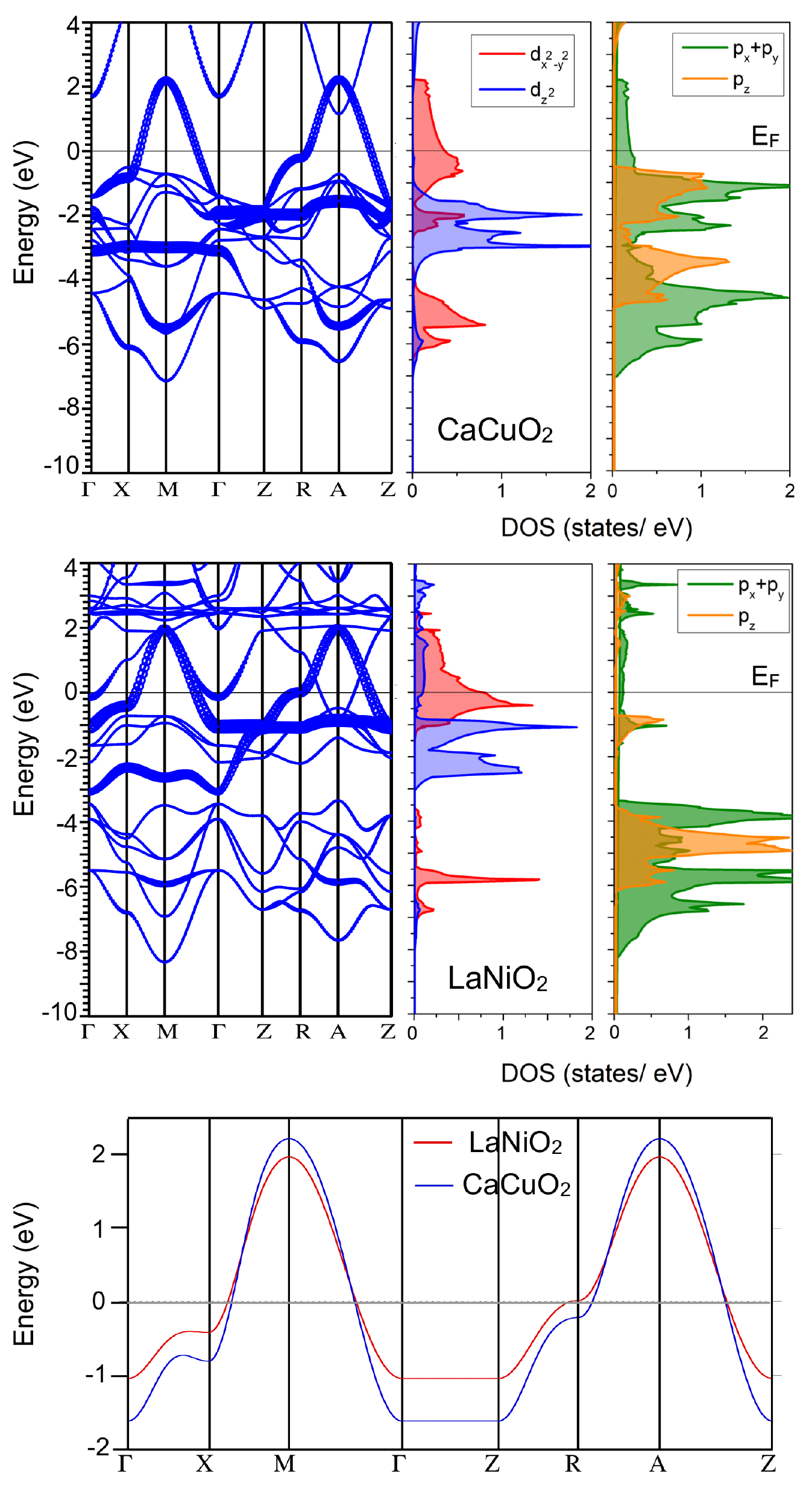}
\caption{Top and middle panels. Comparison of the band structure ($d_{x^2-y^2}$ and $d_{z^2}$ characters highlighted) and orbital-resolved density of states (Ni/Cu-$d_{x^2-y^2}$ and $d_{z^2}$, O-$p_{x}, p_y$ and $p_z$) of CaCuO$_2$ and LaNiO$_2$. Bottom panel. Tight binding fit to the $d_{x^2-y^2}$ band at the Fermi energy for both materials.}
\label{fig1}
\end{figure}

\textit{Comparison of the non-magnetic electronic structures of CaCuO$_2$ and LaNiO$_2$.} Fig.~\ref{fig1} shows the band structures of CaCuO$_2$ and LaNiO$_2$.  To avoid complications connected with the 4$f$ states, we chose to focus on LaNiO$_2$ rather than its Nd counterpart, though we note that the band structure of NdNiO$_2$ is almost identical to that of LaNiO$_2$ as shown in Fig.~2 of Ref.~\cite{supplementary}. Fig.~\ref{fig1} also shows  the orbital-resolved density of states highlighting the Ni/Cu-$d_{x^2-y^2}$ and $d_{z^2}$ and O-$p$ characters for CaCuO$_2$ and LaNiO$_2$. 
As described in previous work \cite{pickett,liu,antia-Pd}, there are some differences between the electronic structures (and magnetic properties, see below) of these two materials. These arise mostly from the different energies of the spacer cation bands.  The Ca-3$d$
bands extend down to about 2 eV above the Fermi level, whereas the La-5$d$ bands dip down and actually cross the Fermi energy, with the pocket at $\Gamma$ having mostly La-$d_{z^2}$ character, that at A La-$d_{xy}$ character.  
These two small electron pockets in the Ni case lead to self-doping
of the large hole-like $d_{x^2-y^2} - p\sigma$ antibonding Fermi surface (Fig.~3 of Ref.~\cite{supplementary}).

It has been suggested that the T$_c$ of the cuprates is correlated with the splitting of the $d_{x^2-y^2}$ and $d_{z^2}$ energies, with a larger value giving rise to a higher T$_c$ due to reduced mixing of these orbitals \cite{sakakibara}. We have compared this energy difference in the Cu and Ni cases using the band centroids calculated as $E_i = \frac{\int g_i(E)EdE}{\int g_i(E)}$, as done in previous work \cite{gw}.  Here, $g_i$ is the partial density of states associated with orbital $i$. The integration range  covers the antibonding band complex for Ni/Cu-e$_g$ states, as in Ref.~\cite{gw}.
The values we derived for CaCuO$_2$ are E$_{x^2-y^2}$= -0.22 eV, E$_{z^2}$= -2.36 eV, giving a splitting of 2.14 eV, consistent with Ref.~\onlinecite{gw}. For LaNiO$_2$, E$_{x^2-y^2}$= 0.20 eV, E$_{z^2}$= -1.75 eV, with a comparable splitting of 1.95 eV. 
These values, though, are quite different from the more physical ones obtained from the Wannier fits (see below).

Another quantity that has been deemed important for determining T$_c$ in the cuprates is the ratio $t^\prime/t$ that describes the relative strength of longer-range hopping to nearest-neighbor hopping in a one-band model - materials with a
larger ratio have a higher T$_c$ \cite{pavarini}. To estimate this ratio, we performed a six-parameter tight-binding fit to the $d_{x^2-y^2}$ band at the Fermi energy.
Values are listed in Table \ref{table1} along with the associated tight-binding functions, and the resulting band structures are plotted in Fig.~\ref{fig1}.
Note that these fits differ from those of Lee and Pickett \cite{pickett}.  In particular,
we considered longer-range in-plane hoppings, and our interlayer functions also differ, in that they take into account the mixing of relevant `even' (with respect to the diagonal mirror plane) states
(4$s$, $d_{z^2}$, $p_z$) with the `odd' $d_{x^2-y^2}$ state that has a $[\cos(k_xa)-\cos(k_ya)]^2$ dependence.  Then, the $t^\prime/t$ ratio is defined as proposed by
Sakakibara {\it et al.}~\cite{sakakibara} as $(|t_3|+|t_2|)/|t_1|$.  The resulting ratio in both cases is quite large, of order 0.4, comparable to that observed for the highest 
T$_c$ cuprates.
We note, though, that $t$ itself for Ni is 80\% of that for Cu.

\begin{table}
\caption{Tight binding fits for the $d_{x^2-y^2}$ band at the Fermi energy for CaCuO$_2$ and LaNiO$_2$, along with the ratio $t^\prime/t$ defined as $(|t_3|+|t_2|)/|t_1|$, with $\epsilon(k) = \sum_i t_i f_i(k)$. Here, $\rm{wt}(k)$ is  $[\cos(k_xa)-\cos(k_ya)]^2/4$.  Units are meV.  $i$ ranges in the table from 0 to 5, and corresponds
to lattice vectors (0,0,0), (1,0,0),
(1,1,0), (2,0,0), (0,0,1) and (0,0,2).}
\begin{ruledtabular}
\begin{tabular}{lrr}
$f_i$ & $t_i$ (LaNiO$_2$) & $t_i$ (CaCuO$_2$) \\
\hline
$1$ & 249 & 201 \\
$2[\cos(k_xa)+\cos(k_ya)]$ & -368 & -460 \\
$4\cos(k_xa)\cos(k_ya)$ & 92 & 99 \\
$2[\cos(2k_xa)+\cos(2k_ya)]$ & -43 & -73 \\
$\rm{wt}(k)\cos(k_zc)$ & -248 & -221 \\
$\rm{wt}(k)\cos(2k_zc)$ & 67 & 50 \\
$t^\prime/t$ & 0.37 & 0.37 \\
\end{tabular}
\end{ruledtabular}
\label{table1}
\end{table}

Finally, the difference in on-site $p$ and $d$ energies (i.e., the charge transfer energy) in cuprates has also been correlated with T$_c$, with smaller values promoting
a larger T$_c$ \cite{weber}. In this context,  the degree of hybridization between the $p$ and  $d$ states is reduced in the Ni case with respect to Cu as can be observed from the orbital-resolved density of states (Fig.~\ref{fig1}). Moreover, the difference between these two on-site energies found from the Wannier fits for LaNiO$_2$ (see below) well exceeds that seen in the cuprates \cite{weber}. This had been previously noted for the related bilayer and trilayer materials La$_3$Ni$_2$O$_6$ and La$_4$Ni$_3$O$_8$ \cite{yee}.

\begin{figure}
\includegraphics[width=0.9\columnwidth]{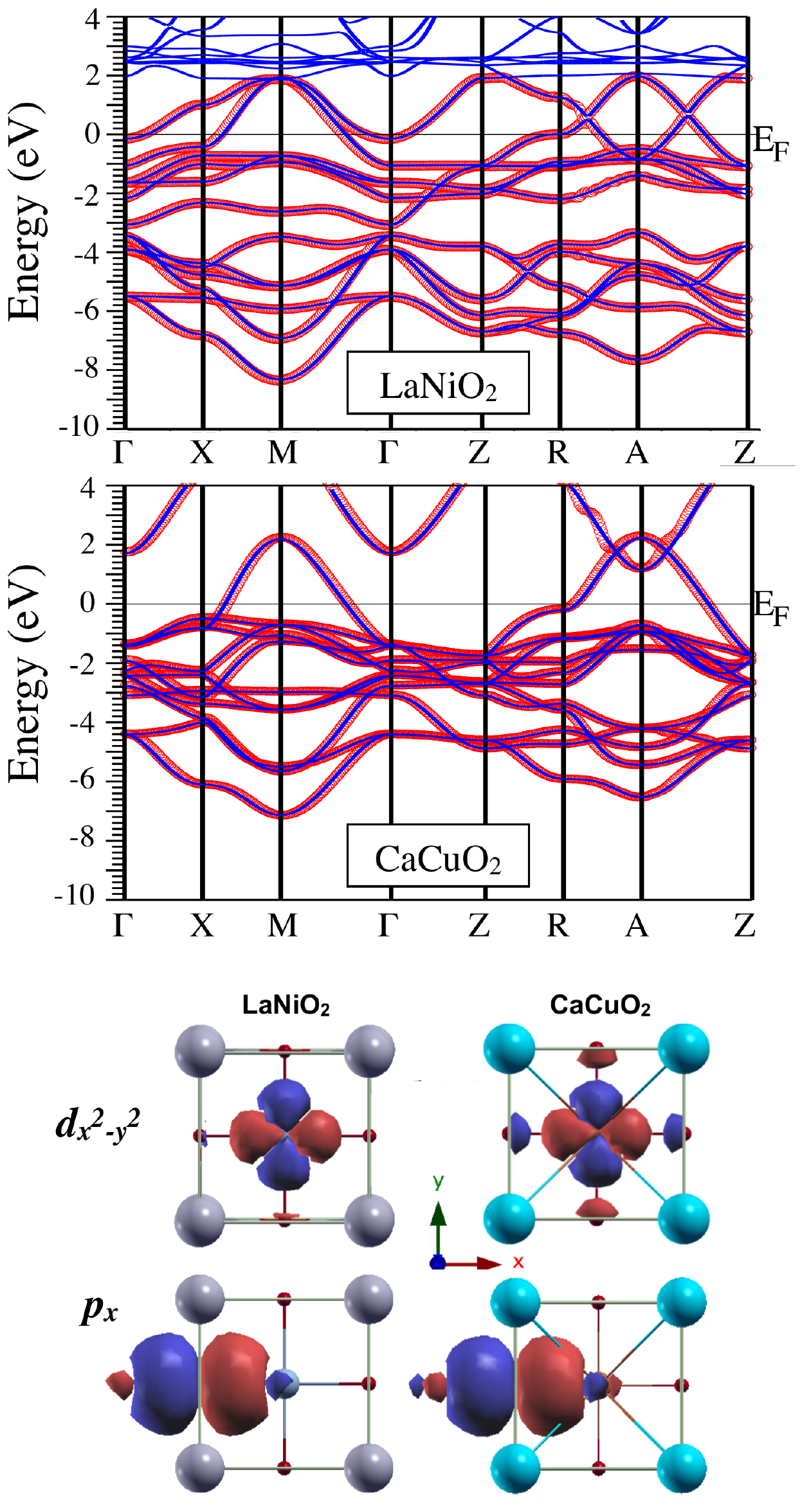}
\caption {Wannier fits (red) and DFT band structures (blue) of CaCuO$_2$ (top) and LaNiO$_2$ (middle).
Comparison of Wannier functions of $d_{x^2-y^2}$ and $p_x$ character for CaCuO$_2$ and LaNiO$_2$ (bottom); the rest are shown in Fig.~4 of Ref.~\cite{supplementary}. Colors represent the sign of the Wannier function.  The large spheres are the Ca (cyan) and La (gray) atoms.}
\label{fig2}
\end{figure}

\textit{Wannierization.} Since the starting
procedure is to assign orbitals localized at specific sites in the
initial projection to obtain MLWFs,
our choice was to take the obvious set of five Cu/Ni-$d$ and six O-$p$ orbitals. These initial associations persisted.
Inclusion of the La/Ca $d_{z^2}$ orbital improves the fits.
Excellent agreement is obtained between the band structures
obtained from the Wannier function interpolation and those derived
from the DFT calculations, showing
a faithful (though not unique) transformation to MLWFs. The Wannier functions describe $d$-like orbitals centered on the Ni/Cu sites (with some O-$p$ contribution for the $d_{x^2-y^2}$ orbitals) and $p$-like on the O sites
(Fig.~\ref{fig2} and Fig.~4 of Ref.~\cite{supplementary}).
The spatial spread \cite{mlwf} of these functions is small and comparable in the Ni and Cu cases ($\sim$1 \AA$^2$).

The on-site energies and hoppings obtained from the Wannier fits are shown in Table \ref{table2}.
The splitting between the d$_{x^2-y^2}$ and d$_{z^2}$ energies (0.7 eV
for Ni, 1.0 eV for Cu) is considerably smaller than that derived from the integration of the density of states mentioned earlier. Moreover, $\Delta$ (referring to $d_{x^2-y^2}$ and $p\sigma$) is 4.4 eV and 2.7 eV for Ni and Cu, respectively. The former, as mentioned above, is well outside the range observed for
cuprates \cite{weber}.
The increased $d$-$p$ splitting in the Ni case leads to a more localized $d_{x^2-y^2}$ Wannier function (Fig.~\ref{fig2})
which could possibly act to promote polaron formation.
Remarkably, the $pd$ and $pp$ hopping parameters are almost identical for the two materials, particularly those relevant for the $d_{x^2-y^2}$ and $p\sigma$ orbitals.

\begin{table}
\caption{Calculated on-site energies and hoppings for CaCuO$_2$ and LaNiO$_2$ derived from the Wannier functions. O1 bonds to Ni/Cu along the $x$ direction,
and O2 bonds to Ni/Cu along the $y$ direction. }
\begin{ruledtabular}
\begin{tabular}{lcc}
\multicolumn{1}{l}{Wannier on-site energies  (eV)} &
\multicolumn{1}{l}{CaCuO$_2$ } &
\multicolumn{1}{c}{LaNiO$_2$   } \\

   \hline
      $d_{xy}$ & -2.55 & -1.75   \\ 
           $d_{xz,yz}$ & -2.44 & -1.65\\ 
              $d_{x^2-y^2}$ & -1.51 & -1.02\\ 
                     $d_{z^2}$ & -2.48 & -1.73\\ 
                     $p_{x}$ O1	& -4.20 & -5.41\\ 
       $p_{y}$ O1 & -2.56 & -4.48\\ 
              $p_{z}$ O1 & -2.72& -4.46\\ 
                 $p_{x}$ O2 & -2.56& -4.47\\ 
                   $p_{y}$ O2 & -4.19& -5.41\\ 
                     $p_{z}$ O2 & -2.72& -4.46\\ 
\hline     
  Wannier hoppings (eV) & & \\ 
\hline        
     $d_{xy}-p_y$ O1 &0.71 & 0.71 \\ 
        $d_{xz}-p_z$ O1 &0.75 & 0.73\\ 
               $d_{x^2-y^2}-p_x$ O1 & -1.20 & -1.23 \\ 
                              $d_{z^2}-p_x$ O1 & 0.25 & 0.20\\

    $p_y$ O2 - $p_x$ O1 &0.53 & 0.59\\ 
        $p_x$ O2 - $p_x$ O1 &-0.33 & -0.27\\ 
          $p_y$ O2 - $p_y$ O1 & 0.33 &0.27 \\ 
                  $p_x$ O2 - $p_y$ O1 & -0.37 & -0.16\\ 
                           $p_z$ O2 - $p_z$ O1 & -0.17 & -0.19 \\
                           
\end{tabular}
\end{ruledtabular}
\label{table2}
\end{table}

We in turn have used these parameters as start values for Slater-Koster fits to the band structures \cite{slater}.  Here, a 19 parameter fit was done using
Powell's method \cite{NR}. These are the
17 parameters used by Mattheiss and Hamann for CaCuO$_2$ \cite{MH} generalized to include two more parameters motivated by the Wannier fits shown
in Table \ref{table2}.  These are
separate $p\pi$ energies for the in plane and out of plane O orbitals, as well as a separate $pd$ hopping integral for $d_{z^2}-p\sigma$.
The rms error we find for the Cu fit is significantly better than that of Ref.~\onlinecite{MH}, and this occurs as well if we restrict to a 17 parameter fit.
This improvement is presumably due to using the Wannier values as start values for the fit.  These fits could presumably be improved if the Wannier
analysis was extended to derive the five interlayer hoppings ($dd\sigma$, $dd\pi$, $dd\delta$, $pp\sigma_{\perp}$, $pp\pi_{\perp}$).
The fit values and resulting band structure are shown in Ref.~\cite{supplementary}.
We find large $pd\sigma$ (1.4-1.5 eV) and $pp\sigma$ (1.2-1.3 eV) values
that are similar between 
Ni and Cu, with a large ratio of $pp\sigma$ to $pd\sigma$ (0.8 for Cu, 0.9 for Ni).
As with the Wannier analysis, the $x^2-y^2$ and $p\sigma$ energies are significantly
different (1.6 eV for Cu compared to 4.3 eV for Ni).

\textit{Spin-polarized calculations of LaNiO$_2$.} A C-type AFM state is the ground state of the system even at the GGA level with magnetic moments inside the Ni spheres of $\sim$ 0.7 $\mu_B$. The FM solution gives rise to a reduced magnetic moment of $\sim$ 0.2 $\mu_B$ at the GGA level, less stable that the C-type AFM state by 0.72 meV/Ni. The energy difference obtained with respect to the non magnetic state is 0.70 meV/Ni and 0.69 meV/Ni with respect to an A-type AFM state. 
The results of the LDA+$U$ calculations reported by Anisimov {\it et al.}~\cite{anisimov} for LaNiO$_2$ gave a stable AFM insulator with the only unoccupied $d$ levels
being those of the minority-spin d$_{x^2-y^2}$ orbital, equivalent to the situation in CaCuO$_2$.
However, the insulating nature of this result could not be reproduced by Lee and Pickett \cite{pickett} and we cannot reproduce it either.
For both the AMF and FLL schemes, the value of the magnetic moment at the Ni site increases up to the highest $U$ value used of 6 eV. 
As noted above, there is no experimental evidence for antiferromagnetic order in LaNiO$_2$. In
fact, the susceptibility looks Pauli-like except for a low
T upturn that is probably due to nickel metal impurities \cite{hayward}. We also note that once $U$ increases
 (U $\geq$ 4 eV), hybridization with the La-$d$ states causes the Ni-$d_{z^2}$ orbitals to rise in energy as reported by Lee and Pickett \cite{pickett}.
 This can be clearly observed in Fig.~6 of Ref.~\cite{supplementary}.

\textit{Doping studies.} One can estimate the effective doping level of the large hole-like Fermi surface versus the actual hole doping using a rigid band approximation.
Because of the self-doping from the two La-$d$ electron pockets at $\Gamma$ and $A$, the doping for optimal T$_c$ could be lower than that observed in most cuprates.
From our rigid band analysis, we estimate this to be 12\% (assuming that optimal doping for the large Fermi surface is at 16\% doping).

\begin{figure}
\includegraphics[width=\columnwidth]{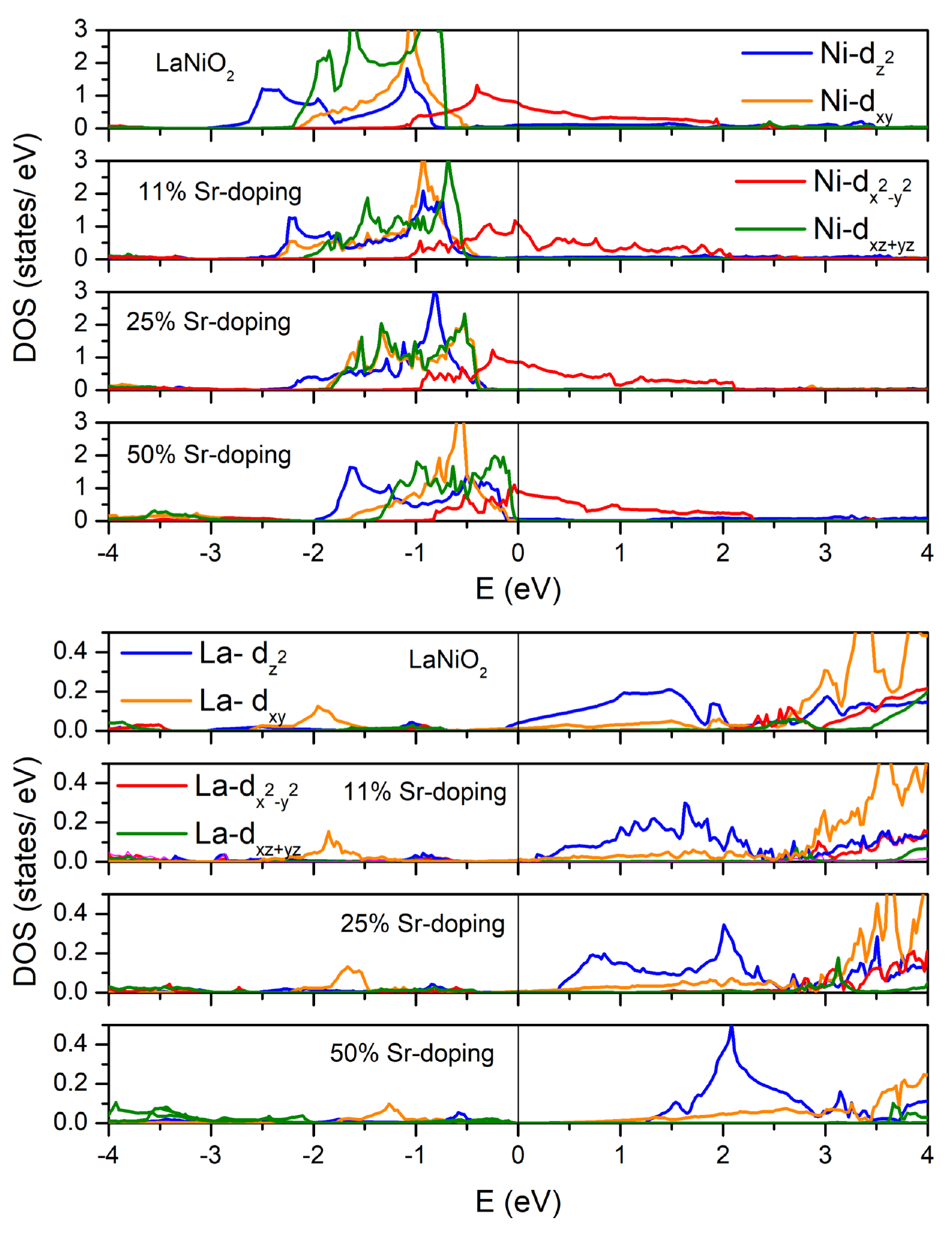}
\caption {Comparison of the orbital-resolved Ni-$d$ (top panels) and La-$d$ (bottom panels) density of states for increasing Sr-doping in LaNiO$_2$.}
\label{fig6}
\end{figure}

For a more sophisticated analysis, we examined the effect of Sr doping by employing supercells that would give rise to an average $d$ filling of 8.89, 8.75, 8.5, respectively. The corresponding orbital-resolved densities of states for La-$d$ and Ni-$d$ are shown in Fig.~\ref{fig6}. The La-$d_{xy}$ character at E$_F$ is diminished upon doping as is the La-$d_{z^2}$ one. The concomitant effect can be seen in the orbital-resolved Ni-$d$ DOS, where, upon doping, the $d_{z^2}$ DOS above E$_F$ is completely suppressed.  This will act to reduce the self-doping effect mentioned above and give a more pure single-band cuprate-like picture.

We now comment on the Hall data.  For zero doping, values of $R_H$ near zero T of -4.6 \cite{ikeda2} and -7.0 cm$^{-3}$/C \cite{ikeda,new} have been reported.  This is inconsistent
with the presence of a large hole-like Fermi surface.  Evaluating
$R_H$ \cite{yrs} using the paramagnetic band structure,
we find a value of -5.2 when restricting to the two small
electron pockets at $\Gamma$ and $A$ (Fig. 3 of Ref.~\cite{supplementary}), as compared to a value
of +0.2 if the entire Fermi surface is included, indeed implying
that the large Fermi surface is gapped out.  For the doped
case, a low T value of +0.4 is reported \cite{new}.  For this doping (0.2), we find using a rigid band shift of $E_F$ a value of +0.2 which rises to +0.3 if the two small pockets are not
included, implying that the large Fermi surface dominates $R_H$ at this doping, consistent with what is found in the cuprates.

To summarize, despite negative conclusions from earlier work, including our own \cite{pickett,antia-Pd}, we find that RNiO$_2$ are promising as cuprate analogs.  Besides the
much larger $d$-$p$ energy splitting, and the presence of 5$d$ states near the Fermi energy, all other electronic structure parameters
seem to be favorable in the context of superconductivity as inferred from the cuprates.  In particular, the large value of $t^\prime/t$ is most promising.  This is not only from an empirical
perspective \cite{pavarini}, such larger ratios also promote longer range exchange \cite{peng}.  In retrospect, then, the discovery of superconductivity in
Sr-doped NdNiO$_2$ should not be as surprising as it has turned out to be.  In that context, we note that electron doping of SrCuO$_2$ leads to
the highest T$_c$ among electron-doped cuprates \cite{smith}.  Therefore, electron doping of RNiO$_2$ could be equally revealing.

The authors would like to thank John Mitchell for discussions.
This work was supported by the Materials Sciences and Engineering
Division, Basic Energy Sciences, Office of Science, US DOE. ASB thanks ASU for startup funds,
and also ANL for hosting her visit where this project was initiated.
We acknowledge the computing resources provided on Blues, a high-performance computing clusters operated by the Laboratory
Computing Resource Center at Argonne National Laboratory.

\end{document}


\subsection{Supplemental Material to ``Similarities and differences between infinite-layer nickelates and cuprates and implications for superconductivity"}

\begin{figure}[H]
\begin{center}
\includegraphics[width=0.4\columnwidth]{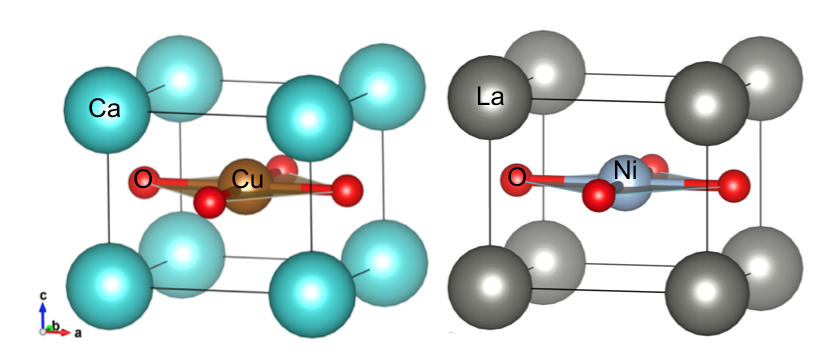}
\caption {Crystal structure of CaCuO$_2$ (left) and LaNiO$_2$ (right).}
\label{fig1}
\end{center}
\end{figure}

\begin{figure}[H]
\begin{center}
\includegraphics[width=0.28\columnwidth]{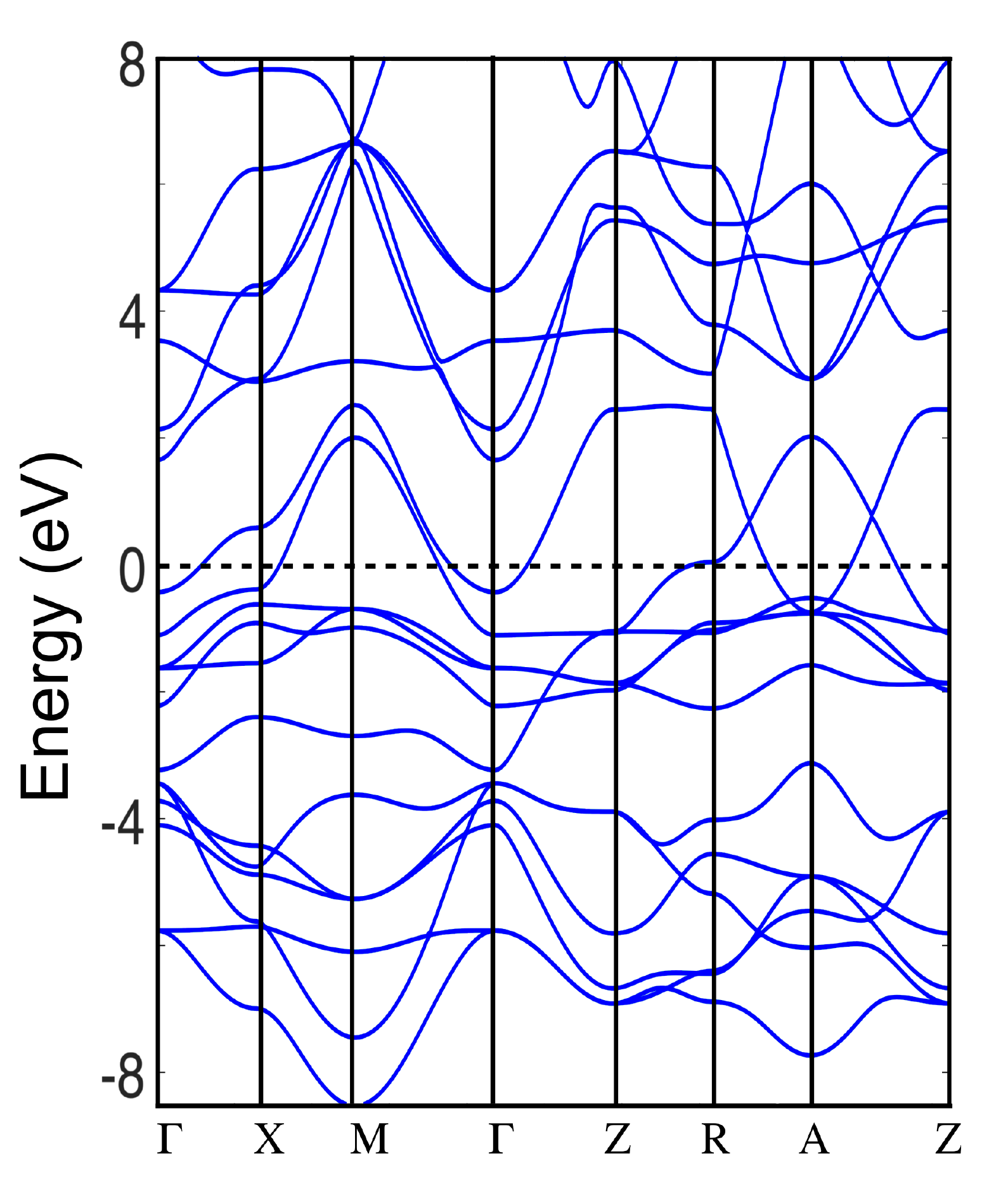}
\caption {Non-magnetic band structure of NdNiO$_2$ with the Nd-4$f$-states in the core. Note that it is almost identical around the Fermi energy to the band structure of LaNiO$_2$ reported in Fig.~1 of the main text.}
\label{fig2}
\end{center}
\end{figure}

\begin{figure}[H]
\begin{center}
\includegraphics[width=0.6\columnwidth]{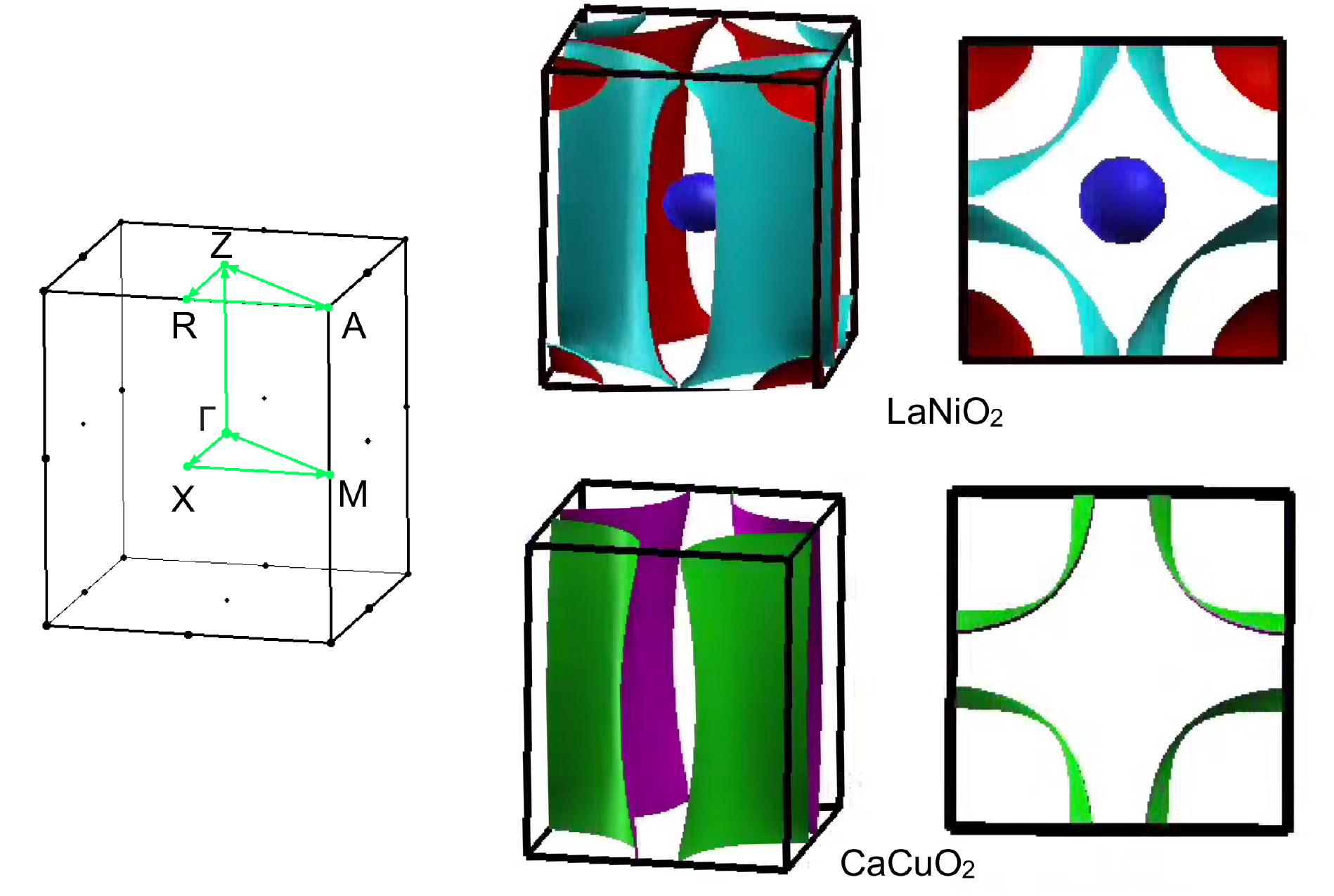}
\caption {Fermi surfaces and Brillouin zone with high symmetry points for LaNiO$_2$ and CaCuO$_2$.}
\label{fig3}
\end{center}
\end{figure}

\begin{figure}[H]
\begin{center}
\includegraphics[width=\columnwidth]{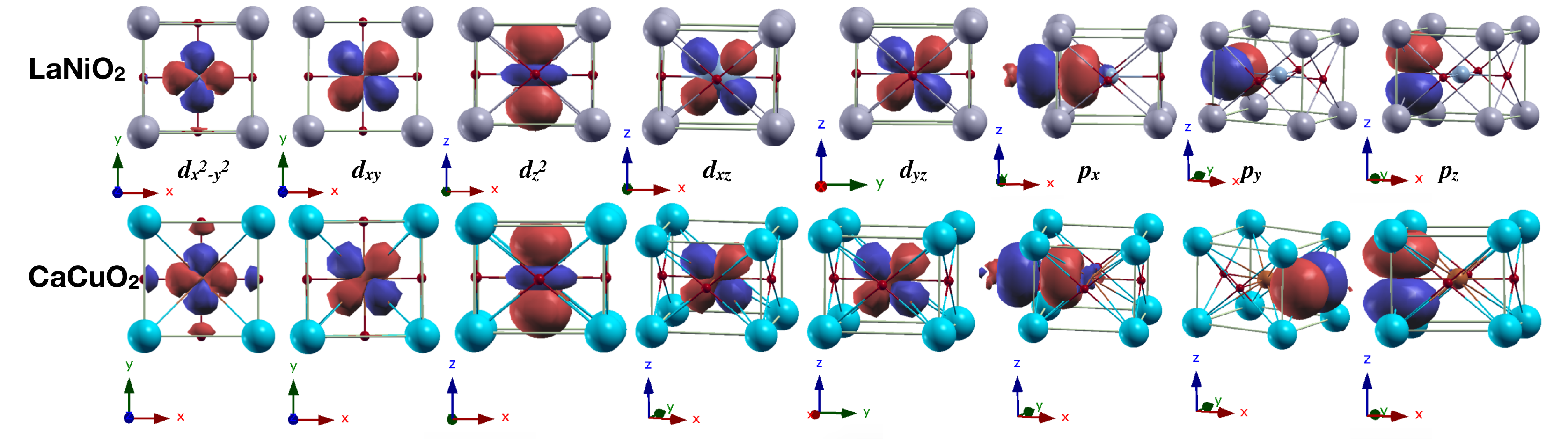}
\caption{3$d$-like (Cu/Ni) and 2$p$-like (O) Wannier functions for LaNiO$_2$ and CaCuO$_2$.}
\label{fig4}
\end{center}
\end{figure}

\begin{table}[H]
\caption{Slater-Koster parameters (in eV) for LaNiO$_2$ and CaCuO$_2$.  $(pp\sigma)_s$ refers to the lobes of the planar O $p$ orbitals pointing towards Ni/Cu, $(pp\sigma)_p$
refers to the alternate configuration where they are each rotated in plane by 90$^\circ$, with $pp\sigma$ referring to when the two are parallel instead (as described by Mattheiss and Hamann \cite{MH}).
Note that the last five values in the table refer to hopping between the planes.
When comparing to the Wannier values  shown in the main text, there are numerical coefficients involved.  For instance, the $d_{x^2-y^2}-p_x$ Wannier term should be compared
to $\frac{\sqrt{3}}{2}pd\sigma$ as tabled in Slater and Koster \cite{slater}.
For the fits, the rms error for Cu is 174 meV and for Ni 331 meV.
The larger value for Ni is likely due to not taking into account the influence of the La $d$ states.}
\begin{ruledtabular}

\begin{tabular}{lrrrr}
  & LaNiO$_2$ & CaCuO$_2$ & & \\
 \hline
$E_{xy}$ & -1.94 & -2.54 & & \\
$E_{xz}$ & -1.72 & -2.33 & & \\
$E_{z^2}$ & -1.84 & -2.57 & & \\
$E_{x^2-y^2}$ & -1.05 & -2.17 & & \\
$E_{p\sigma}$ & -5.39 & -3.72 & & \\
$E_{p\pi}$ & -4.39 & -2.58 & & \\
$E_{pz}$ & -4.58 & -2.85 & & \\
$pd\sigma$ & -1.41 & -1.48 & & \\
$pd\sigma-z^2$ & -0.54 & 0.86 & & \\
$pd\pi$ & 0.79 & 0.74 & & \\
$pp\sigma$ & 0.60 & 0.88 & & \\
$(pp\sigma)_s$ & 1.31 & 1.21 & & \\
$(pp\sigma)_p$  & 0.40 & 0.61 & & \\
$pp\pi$ & -0.05 & -0.20 & & \\
$dd\sigma$ & -0.43 & -0.33 & & \\
$dd\pi$ & 0.08 & 0.02 & & \\
$dd\delta$ & -0.09 & -0.02 & & \\
$(p\sigma)_{\perp}$ & 0.19 & 0.38 & & \\
$(p\pi)_{\perp}$ & 0.02 & -0.01 & & \\
 \end{tabular}
 \end{ruledtabular}
\label{table1}
\end{table}

\begin{figure}[H]
\begin{center}
\includegraphics[width=0.48\columnwidth]{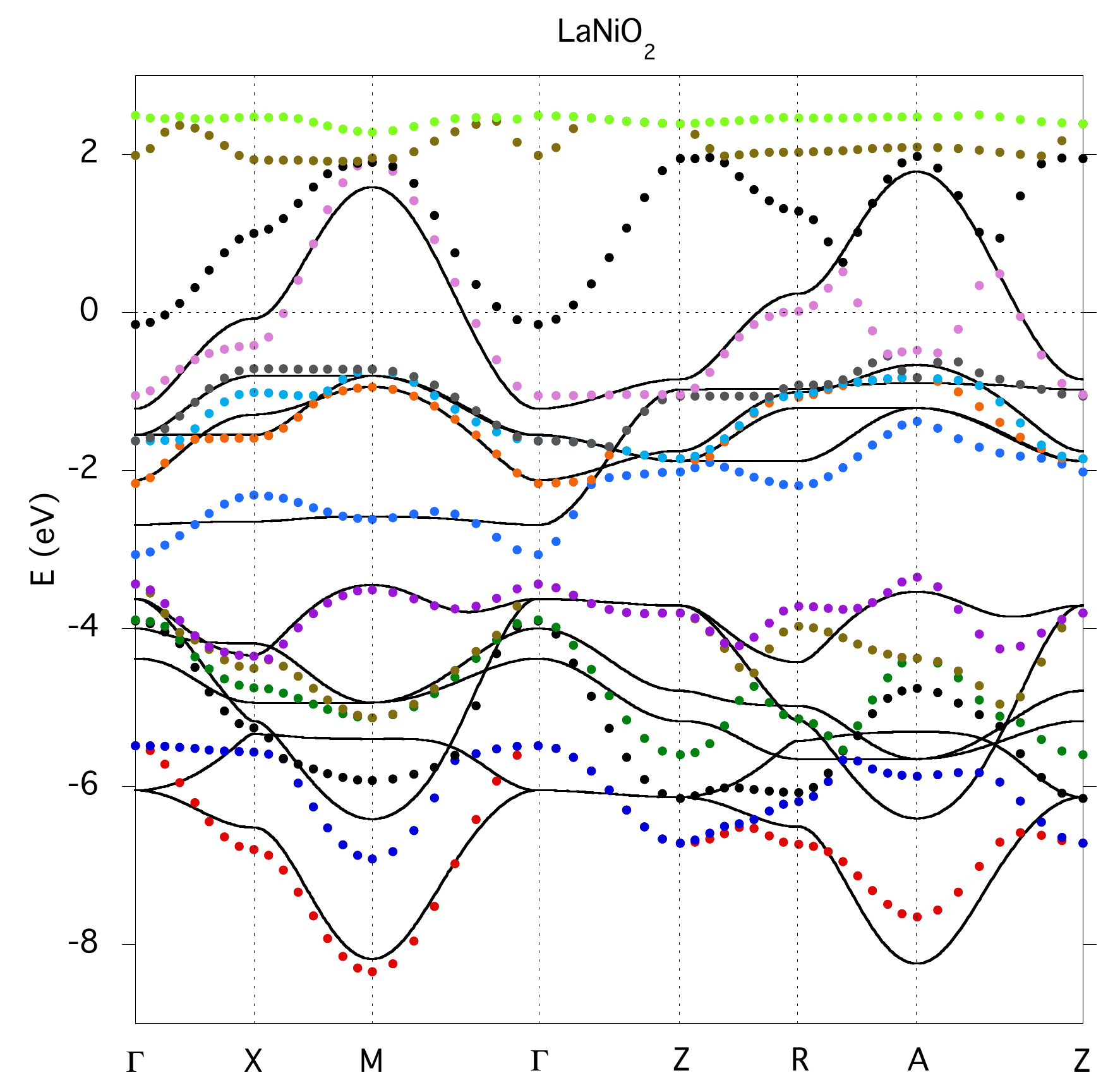}
\includegraphics[width=0.48\columnwidth]{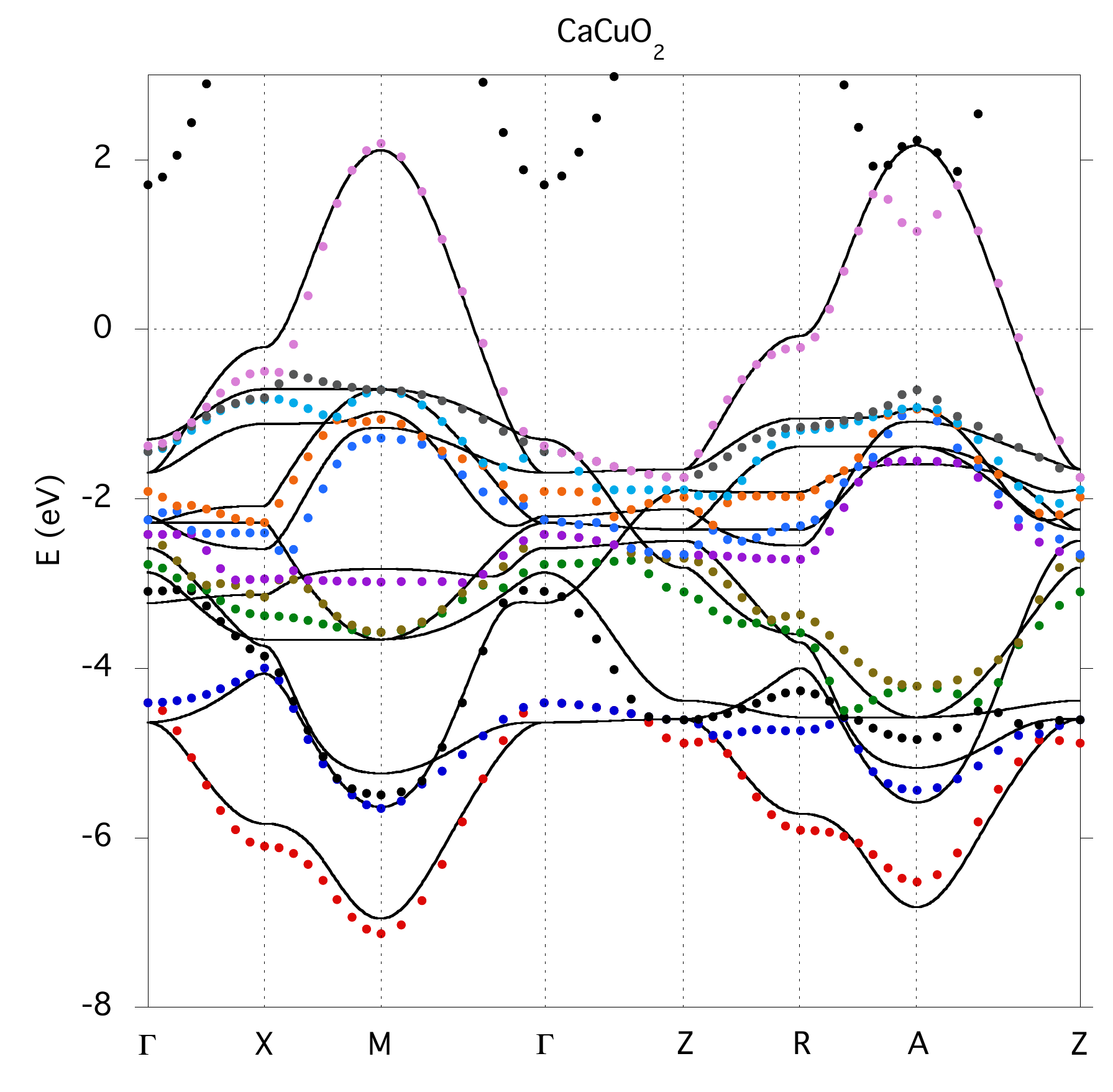}
\caption {Slater-Koster fits for LaNiO$_2$ and CaCuO$_2$.  The solid points are from the GGA
band structures.}
\label{fig4}
\end{center}
\end{figure}

\begin{figure}[H]
\begin{center}
\includegraphics[width=0.46\columnwidth]{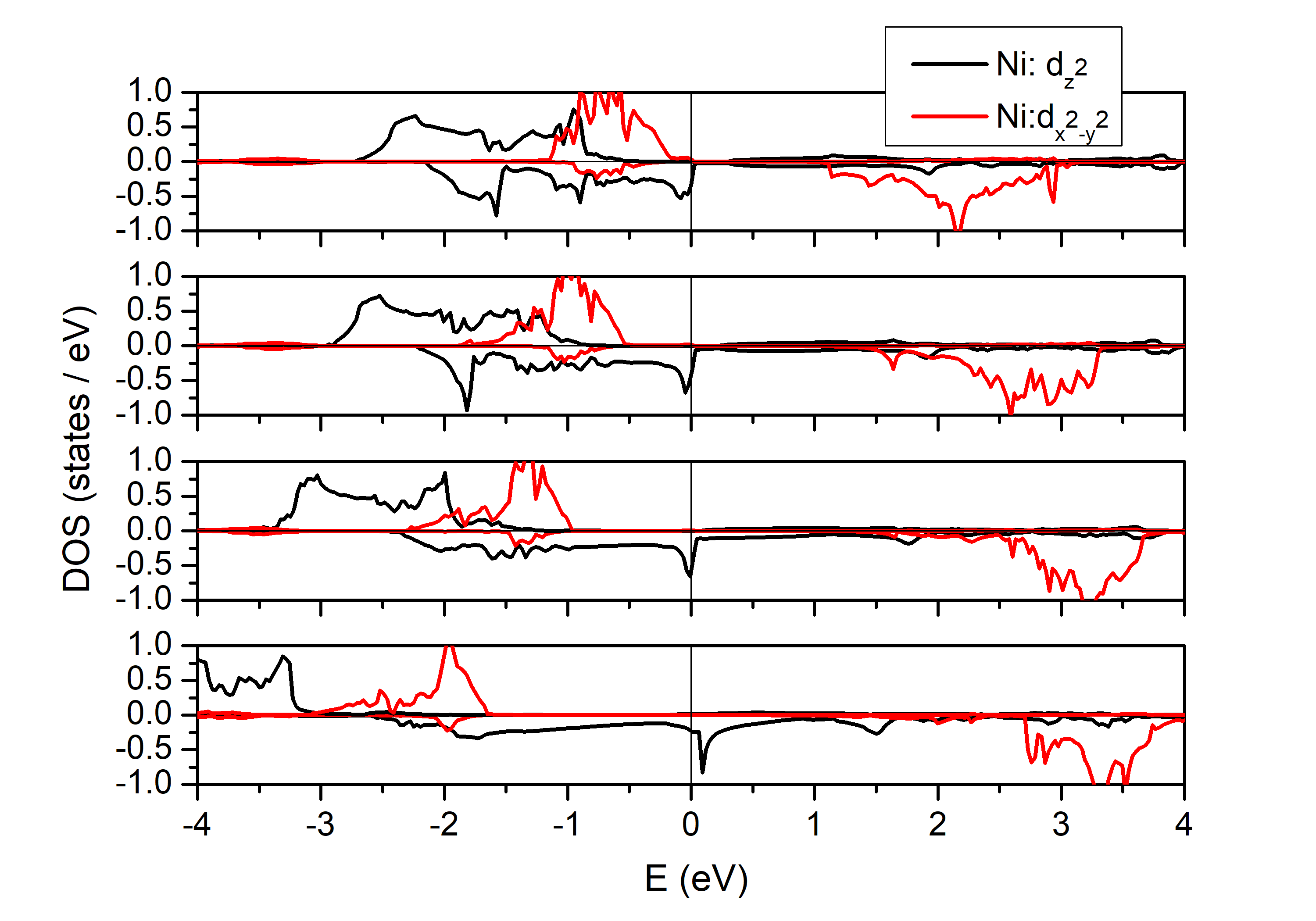}
\includegraphics[width=0.46\columnwidth]{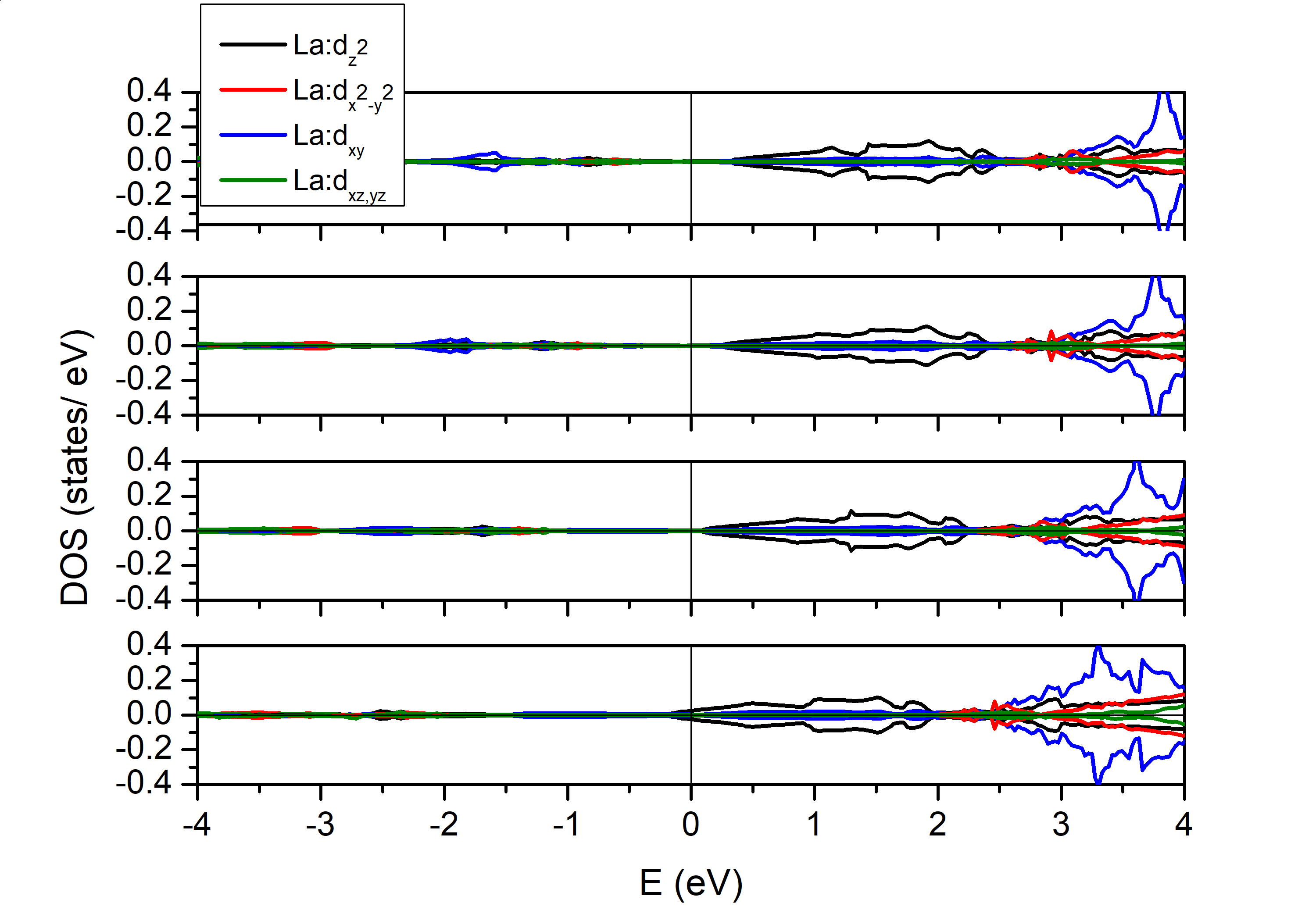}
\caption {Comparison of the orbital-resolved Ni-$d$ (left panels) and La-$d$ (right panels) density of states for a C-type AFM state in LaNiO$_2$ with increasing $U$ ($U$=1.4 eV, 2.7 eV, 4 eV, 6 eV from top to bottom).}
\label{fig3}
\end{center}
\end{figure}